\documentclass[usenatbib]{aa}

\usepackage{graphicx}
\usepackage{amssymb}
\usepackage{epsfig}
\usepackage{natbib}
\bibpunct{(}{)}{;}{a}{}{,}

\def\B{{\em BeppoSAX }}

\begin{document}

\title{The broad-band spectrum of Cyg X-2 with INTEGRAL}
\author{G. Lavagetto\inst{1}\fnmsep\thanks{\email{lavaget@fisica.unipa.it}},
T. Di Salvo\inst{1}, M. Falanga\inst{2},R. Iaria\inst{1},
N.R. Robba\inst{1}, L. Burderi\inst{3}, W.H.G. Lewin\inst{4}, M. M\'endez\inst{5},
L. Stella\inst{6}, and M. van der Klis\inst{7}}

\offprints{G. Lavagetto}
\titlerunning{Spectral Analysis of Cyg X-2 with INTEGRAL}
\authorrunning{G. Lavagetto et al.}
\institute{Dipartimento di Scienze Fisiche ed Astronomiche,
  Universit\`a di Palermo, via Archirafi 36, 90123 Palermo (PA) Italy
\and CEA Saclay, DSM/DAPNIA/Service d'Astrophysique (CNRS FRE
  2591), F-91191, Gif sur Yvette, France
\and Universit\`a degli Studi di Cagliari, Dipartimento di Fisica, SP 
Monserrato-Sestu, KM 0.7, 09042 Monserrato, Italy
\and Center for Space Research, Massachusetts Institute of Technology,
77 Massachusetts Avenue, Cambridge, MA 02139 
\and SRON, National Institute for Space Research, 3584 CA, Utrecht,
the Netherlands
\and Osservatorio Astronomico di Roma, Via Frascati 33, 
00040 Monteporzio Catone (Roma), Italy
\and Astronomical Institute 'Anton Pannekoek', University of
Amsterdam, and Center for High Energy  Astrophysics, Kruislaan 403,
1098 SJ, Amsterdam, the Netherlands
 }

\abstract{We study the broad band (3-100 keV) spectrum of Cygnus X--2
  with {\em INTEGRAL}. We find that the spectrum is well fitted by a
  Comptonized component with a seed-photons
  temperature of $\sim 1$ keV, an electron temperature of
  $\sim 3$ keV and an optical depth $\tau \sim 8$. Assuming spherical
  geometry, the radius of the seed-photons emitting region is $\sim
  17$ km.  The source shows no hard X-ray emission; it was detected
  only  at a $3 \sigma$ level above 40 keV. We also analyzed public
  ISGRI data of Cyg~X--2 to investigate the presence of a hard
  X--ray component. We report the possible presence of hard X-ray emission  
 in one data set.

\keywords{accretion, accretion disks -- binaries: close -- stars:
individual: Cygnus X--2 -- stars: neutron} }
\maketitle

\section{Introduction}
\label{sec:intro}
Cyg~X--2 is one of the six Galactic Low Mass X-ray Binaries (hereafter
LMXBs) that are classified as Z-sources. This classification relies
upon a combination of the track (that resembles a Z) traced on an X-ray
color-color diagram (CCD) and of the correlated timing properties
of these sources (Hasinger \& van der Klis 1989).

Several studies have been carried out on Cyg~X-2:
type-I X-ray bursts have been observed in its X-ray lightcurve (Smale 1998), confirming that the primary  in the system is a neutron star (NS);
 its distance has
been estimated to be $7.2 \pm 1.1$ kpc from
optical observations (Orosz \& Kuulkers 1999), consistent with previous
determinations from radio observations (Hjellming et al. 1990) but not with measurements obtained from X-ray bursts (11.6 kpc, Smale 1998). The
binary system has an orbital period of 9.8444 d, as can be deduced
from the optical behavior of the companion, V1341~Cyg (Casares et al. 1998).
The masses of the two stars are $1.78 \pm 0.23 M_\odot$  and $0.60 \pm 0.13 M_\odot$ for the primary
and the companion, respectively (Orosz \& Kuulkers 1999).


The X-ray spectrum of the source has been studied several times over
the years.  It has
been fitted both  with the so-called Western model -- a blackbody plus a
Comptonized component (Hasinger et al. 1990;  Smale et al. 1993), and with
the so-called Eastern model -- a multi-temperature blackbody together
with a Comptonized blackbody (Hasinger et al. 1990; Hoshi \& Mitsuda
1991; Hirano et al. 1995). More recently, the broad band spectrum of
Cyg~X--2 has been studied with \B (Frontera et al. 1999; Di Salvo et
al. 2002; Piraino et al. 2002).

Di Salvo et al. (2002) fitted the broad band spectrum of Cyg~X-2
obtained with \B\, using a two component continuum model,
consisting of a disk blackbody and a Comptonized component. A broad emission line at $\sim 1$ keV and an emission line at
$\sim 6.7$ keV (most probably coming from highly ionized
iron) proved necessary for a good fit. In two of the intervals selected on the CCD, the continuum spectrum
of Cyg~X--2 could not be fit by the usual two-component model, and a
third component was needed to fit the high energy part of the
spectra. This hard
X-ray emission fits a power-law with photon index $\sim 2$.

A similar ``hard X-ray tail'' has been found in other observations,
but with an index  $\Gamma \sim 3$ (Piraino et al. 2002).

A hard X-ray emission has been detected  in other Z-sources
:   
GX 5--1 (Asai et al. 1994),  GX~17+2 (Di Salvo et al. 2000),
 GX~349+2 (Di Salvo et al. 2001), Sco X--1 (D'Amico 
  et al. 2001a), and in the 
non-flaring state of the anomalous Z-source Cir X--1 (Iaria et
al. 2001). The physical mechanism producing this high-energy
emission is yet to be understood.
Some evidence of the correlation of the
hard tail with the position of the source in the X-ray CCD has been
found for GX 5--1, GX~17+2 and Cyg X--2. Such a correlation may not be
present in observations of  Sco X--1 (D'Amico 
et al. 2001a) and of GX~349+2 (see Iaria et al. 2004).
D'Amico (2001b) proposed that the presence of a
high-energy non-thermal emission could be linked to the 20--50 keV
luminosity of the Comptonized component of the spectrum: they made the hypothesis
that there is a threshold luminosity ($\sim 4 \times
10^{36}~\mathrm{ergs~s^{-1}}$) of the
Comptonized component above which the non-thermal hard X-ray emission
is  produced. 

In analyzing the first {\em INTEGRAL} pointings to Cyg~X--2, Natalucci
et al. (2003) found the source in the ISGRI images in the 20-40 keV energy
band, while it was not detected above 40 keV.
We report here on the broad band (3--100 keV) spectral analysis of an
{\em INTEGRAL} AO1 observation of Cyg~X--2 and of all the
publicly available ISGRI observations of the source.

\section{Observations}

 The AO1 observation was performed between
May 2 and May 3, 2003 (52761.32-52762.43 MJD), during satellite
revolution 67. The high-energy coded mask imager 
IBIS/ISGRI (Ubertini et al. 2003; Lebrun et al. 2003)  was used to observe Cyg~X-2 for a total exposure
time of 66.8 ks. Cyg X-2 was in the JEM-X (Lund et
al. 2003) field of view during four science windows, for a total exposure
of  7.2 ks. Data were extracted for all pointings with a
source position offset $\leq$ $12^{\circ}$ in ISGRI and $\leq$
$3.5^{\circ}$ in JEM-X. 

Publicly available data from the satellite pointings towards the
Cygnus region during the Performance Verification Phase (PVP) have been previously analyzed
(Natalucci et al. 2003). Along with these data, 10 more observations have been
carried out with the high-energy coded mask imager and have become
publicly available. The publicly available observations used in this
paper are listed in Table 1.
 
\begin{table}[ht]
\begin{center}
\begin{tabular}{lllllll}
\hline
 \hline         
Obs. &Start &  End & Exposure (s) & Rev & P.L.& ASM \\
& & & & & norm. & rate\\
\hline
1&1085.63  & 1087.82  & 20737    & 023 & $<9.1$ & $29 \pm 1.5$\\
2&1092.68  & 1093.51  & 9169.59  & 025 & $<9.0$ & $54.9 \pm 0.2$\\
3&1109.35  & 1109.39  & 2209     & 031 & $16.8 \pm 8.4 $& $34.4 \pm 0.5$\\
4&1178.89  & 1178.92  & 3308.26  & 054 & $<12.7$  & $31.0 \pm 0.6$\\
5&1202.11  & 1202.17  & 4992.27  & 062 & $<9.2$ & $32.4 \pm 0.6$\\
6&1226.06  & 1226.12  & 4917.2   & 070 & $<9.1$ & $40.5\pm 0.3$\\
7&1238.05  & 1238.1   & 4889.13  & 074 & $<9.2$ & $39.4 \pm 0.4$\\
8&1257.39  & 1258.34  & 17113.3  & 080 & $<9.2$ & $41.4 \pm 0.6$\\
9&1261.95  & 1262     & 4913.84  & 082 & $<9.2$ & $38.9 \pm 0.9$\\
10&1291.96  & 1292.96  & 5021.14  & 092 & $<7.8$ & $41.0 \pm 1.2$\\
11&1441.42  & 1441.45  & 3308.99  & 142 & $<7.1$ & $35.2 \pm 0.4$\\
12&1450.43  & 1450.48  & 4923.44  & 145 & $<4.0$ & $31.8 \pm 0.6$\\
AO1&1217.37 & 1218.42 & 66795.96 & 067 & $<4.0$ & $45 \pm 3$\\
\hline
\end{tabular}

\caption{ Summary of all ISGRI observations
  of Cyg~X--2 used in this paper. Start and stop times are in IJD
  (IJD= MJD  - 51544). The 
criterion used to detect the hard X-ray emission is
discussed in section \ref{sec:ht}.Here we report the power--law flux
in the energy range 30--100 keV in units of $10^{-11}$ erg
cm$^{-2}$ s$^{-1}$. Errors and upper limits are at 90\% confidence.} 
\end{center}
\label{ostab}
\end{table}

Data reduction for both instruments was performed using the standard Offline Science
Analysis (OSA) version 4.2 distributed by the {\it INTEGRAL} Science
Data Center (Courvoisier et al. 2003). The algorithms used in the analysis
are described in Goldwurm et al. (2003).
For the spectral analysis we used data corresponding to the 3--20 keV
energy range for JEM-X and to the 20--100 keV energy range for ISGRI. 

\section{Results}
\label{sec:res}

\subsection{The AO1 observation}

\begin{figure}
\centerline{\epsfig{file=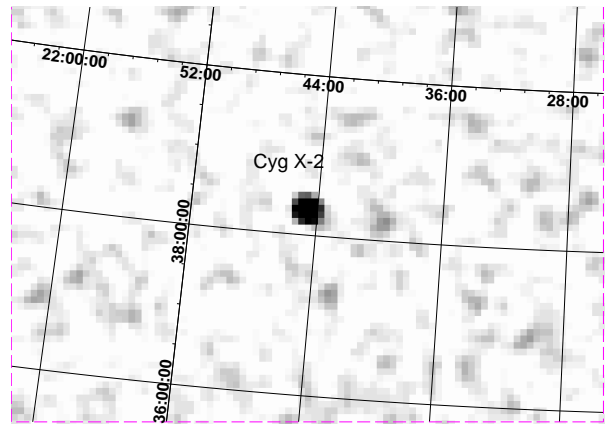,width=8cm}}
\caption
{
 The 20--40 keV IBIS/ISGRI mosaicked and deconvolved sky image of the
AO1 66.8 ks observation.
The image is centered at Cyg X-2
position. The pixel size is 5$'$.
}
\label{fig:ibis_img}
\end{figure}

Fig. \ref{fig:ibis_img} shows the ISGRI map of the Cyg X-2 region
in the 20--40 keV energy range during the AO1 observation. Single pointings were 
deconvolved and analyzed separately, and then combined in mosaic images.
The source is clearly detected at a significance level of
$36\sigma$. In the energy range 40--80 keV, the source can be detected
only with a significance of $3\sigma$. At higher energies Cyg X-2 was not detected at a statistically
significant level neither in single exposures nor in the total exposure time.
Cyg~X-2 was observed in ISGRI images at $\alpha_{\rm J2000} = 21^{\rm h}44^{\rm m}40^{\rm s}$ and $\delta_{\rm J2000} =
38{\degr}19\arcmin00\arcsec$. The source position is consistent,
within the source location error at 90\% confidence  of $0.8\arcmin$ (see Gros et al. 2003),
with the catalog position (Liu et al. 2001).

\begin{figure}[h]
  \centering
  \epsfig{file=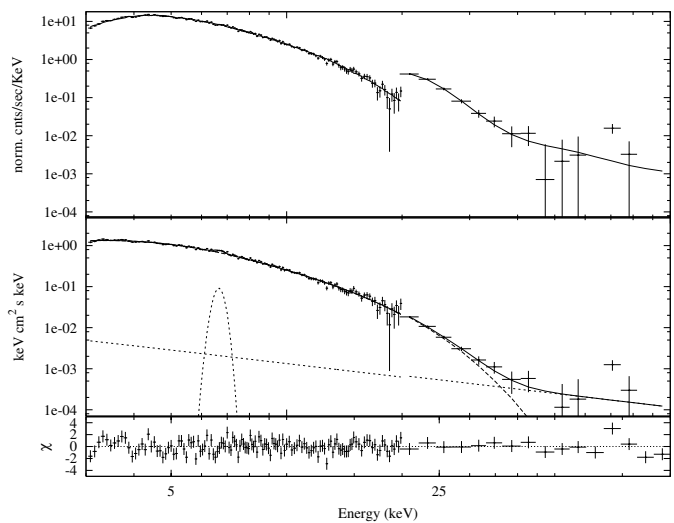,height=10cm, angle=270}
  \caption{Folded (top panel) and unfolded (middle panel) spectra of the AO1 Observation of Cyg~X--2 (top panel) together with residuals with respect to the best fit model (bottom panel).}
  \label{fig:eufde}
\end{figure}

We base our spectral investigations mainly on preceding \B\, spectral results (Di Salvo et al. 2002).
We find that the broad-band (3--100 keV) spectrum of the AO1
observation is well fitted by an absorbed Comptonization spectrum. Due
to the lack of low-energy data we fixed the value of the
photoelectric  absorption to the value reported by Di Salvo et
al. (2002). We will adopt their spectral model in fitting the data.
Using the \textsc{compTT} (Titarchuk 1994) model in xspec 11.3.1 we were able
to fit the data with a reduced chi squared of 1.09 for 144 degrees of freedom
(d.o.f.). The non-detection of a low energy thermal component in the continuum
spectrum  is not
surprising, since it was used to model mainly the low energy (0.1--3
keV) \B\, spectra.  The addition of a power-law component at high
energies, or of a Gaussian emission line at $\sim 6.7$ keV that were
found in \B\, observations are not statistically required.
 Fixing the energy and the width of the emission line to the best
values obtained by Di Salvo et al. (2002), we find an upper limit on
the intensity of $13 \times 10^{-3}$ photons cm$^{-2}$ s$^{-1}$ (at 90
\% confidence), well compatible with the intensity found in the \B\,
observations,  which
varied between 3 and $7 \times 10^{-3}$ photons cm$^{-2}$ s$^{-1}$.
%
%
 We were able to determine an upper limit on the flux of a
power-law component (using the \textsc{pegpwrl}
model inside xspec) by fixing the photon index to the value reported
by Di Salvo et al. (2002), 2.09.  We find an upper limit (at 90 \%
confidence) on power law flux between 30 and
100 keV of $4.0 \times 10^{-11}$ erg cm$^{-2}$ s$^{-1}$.
During the \B\, observation analyzed by Di Salvo et al. (2002) in
which the hard X-ray emission was detected, 
the flux in the same energy band was $ 5.2^{+7.9}_{-1.1}\times
10^{-11}$  erg cm$^{-2}$ s$^{-1}$\footnote{All errors on the fluxes
  are at 90\% confidence.}. Since the flux in
the 3--30 keV energy band in the present observation, $(1.1 \pm 0.3)
\times 10^{-8}$ erg cm$^{-2}$ s$^{-1}$, is comparable
with the flux in the \B\, data in the same band, $(1.0 \pm 0.3)\times
10^{-8}$   erg cm$^{-2}$ s$^{-1}$, we conclude that the spectrum of
the {\em INTEGRAL} data is softer than during the \B\, observations  and that no significant hard X-ray emission is present. We added a normalization factor between Jem-X and ISGRI spectra; keeping the ISGRI normalization fixed to 1 we found a Jem-X normalization of 0.986.
Detailed results for the best-fit
model are reported in table \ref{tab:data}. The reduced $\chi^2$
(d.o.f.) of the best fit model is 1.03(142).
 The folded and unfolded spectra together with the best-fit model and the respective
residuals are shown in figure \ref{fig:eufde}.

\begin{table}[htbp]
  \centering
  \begin{tabular}{lll}
   \hline
\hline
 & CompTT & CompTT +\\
& & power-law \\
\hline
N$_\mathrm{H}$ ($\times 10^{22}$ cm$^{-2}$) &0.2 (frozen) & 0.2 (frozen) \\
kT$_\mathrm{W}$ (keV) & $0.99\pm 0.03$ & $0.95 \pm 0.04$\\
kT$_\mathrm{e}$ (keV) &$3.44^{+0.43}_{-0.34}$& $3.03^{+0.41}_{-0.31}$\\
$\tau$ & $6.8^{0.87}_{-0.82}$& $8.0^{+1.2}_{-1.1}$\\
R$_\mathrm{W}$&$17.1\pm 1.6 $ & $18.3 \pm 2.2$\\
Pho. Index & - &2.09 (frozen)\\
N$_\mathrm{po}$& - &$ <4.0$\\
E$_{Fe}$(keV) & - & 6.65 (frozen)\\
$\sigma_{Fe}$(keV) & - & 0.2 (frozen)\\
I$_{Fe}$ & - & $<13 \times 10^{-3}$\\
Flux & 1.05 & 1.13\\
$\chi^2$(d.o.f.) &1.09(144)& 1.03(142)\\ 
\hline
  \end{tabular}
  \caption{Results of the fit of the {\em INTEGRAL} Cyg~X-2 spectrum in
    the 3--100 keV energy band for the AO1 observation, with a Comptonized spectrum modeled by
    CompTT and with a model
    consisting of compTT and a power law with the photon index frozen to $2.09$. Uncertainties are at 90\% confidence level for a
    single parameter. Upper limits (also at 90\% confidence) for the normalization of a Gaussian
    emission line from highly ionized iron and of a hard power-law are reported here to make a comparison with
    results by Di Salvo et al. (2002). $kT_W$ is the temperature of
    the soft seed photons for the Comptonization, $kT_e$ is the
    temperature of the scattering electron cloud, while $\tau$ is
    its optical depth in spherical geometry. R$_\mathrm{W}$ is the
    radius of the
    seed photons emitting region, assuming a source distance of 7.2
    kpc. The power-law flux  is in units of
     $10^{-11}$ erg cm$^{-2}$ s$^{-1}$ in the 30--100 keV energy
     range.  The emission line
    intensity I is in units of photons cm$^{-2}$ s$^{-1}$.
 The total
    unabsorbed flux,
    in units of 10$^{-8}$ ergs cm$^{-2}$ s$^{-1}$, refers to the
    3--100 keV energy range. }
  \label{tab:data}
\end{table}

\subsection{Looking for hard X-ray emission in other INTEGRAL
  observations of Cyg~X-2}
\label{sec:ht}
Since no JEM-X data are available during other public ISGRI
observations, it is not possible to determine the parameters of the
{\sc compTT} component at these times. To test the presence of a hard tail, 
we fitted the ISGRI data with the model we fitted to the AO1
observation, keeping the seed photons temperature $T_\mathrm{w}$ and
the optical depth of the Comptonizing cloud $\tau$ frozen to their best-fit values; the only free parameters in
the fit are therefore the normalizations of the two components and the
electron temperature of the {\sc compTT} component. In 
observations 3, 4 and 11 (see table 1) we had to fix the
electron temperature as well.

We consider that a hard tail is present in the spectrum when the power-law
flux in the 30-100 keV energy range is different from 0 at 90\% confidence.
Using this criterion, we find one data set in which the hard tail is
 detected (see table 1). In the other 11 data sets, no hard
 tail is detectable, although in most cases the upper limits are
 consistent with the flux found by \B. The detection (or the lack of it) is reported in table 1.
Our results partially confirm that a hard tail appears from time to
time in  
Cyg~X--2 spectra, as reported by other authors (Di Salvo et al. 2002, Piraino et
al. 2002). It was proposed that the presence of the hard power-law
emission is strongest in the horizontal branch of the X-ray CCD (Di Salvo et al. 2002). Unfortunately we have no possibility
to obtain information about the position of the source along its Z
track, neither in the public data nor during the AO1 observation (JEM-X
data covered only a small fraction of the ISGRI
observation). In the analysis of the first {\em INTEGRAL} observations
of Cyg~X--2, Natalucci et al. (2003) suggested that there
could be a correlation between the variability of the high-energy
spectrum and the long term variability in the soft X-ray lightcurve of
Cyg~X--2. We analyzed the Cyg~X--2 lightcurves obtained with the RXTE
All Sky Monitor (ASM) simultaneously with the {\em INTEGRAL}
observations, seeking for some sort of correlation between the long
term soft X--ray variability and the detection of a hard tail. We
do not find evidence of a correlation between the presence of the hard tail
and the count rate in the ASM (see table 1).

 We tried to obtain a
crude estimate of the  20-50 keV luminosity of the Comptonized component from our
fitting model in order to verify the hypothesis by D'Amico et al
(2001b).  In most observations, the error on the inferred luminosity
is of one order of magnitude.
In the AO1 observation however the power-law was not detected and the
20-50 keV luminosity of the Comptonized component is $(1.0\pm 0.2) 10^{36} ~\mathrm{erg~
  s}^{-1}$, but in observation 3, where we detect the hard tail,
the luminosity of the Comptonized component is $(0.6 \pm 0.5)\times
10^{36}$ erg~s$^{-1}$. This does not agree with the hypothesis of D'Amico et
al. (2001b).

\section{Conclusions}

We analyzed the {\em INTEGRAL} AO1 observation of Cyg~X-2. The source
was  clearly detected in the 20-40 keV (see Figure 1), but only at a 3
$\sigma$ level in the 40-80 keV energy band.
The broad band spectrum is well
fitted by a Comptonization model alone. No hard tail was significantly
detected in the data
We also analyzed 12 publicly available observations carried out with
ISGRI, and confirmed the detection of a hard tail
at 90\% confidence in one case. This is the first direct
detection of a hard tail in high-energy imaging
data of a Z-source. A longer, continuous
observation of Cyg~X--2 with both JEM-X and ISGRI would allow to investigate further the nature
of the hard X-ray emission and its relationship with other spectral features
of the source. 




\acknowledgements
We acknowledge financial contribution from contract ASI-INTEGRAL I/R/046/04

\end{document}